\documentclass[useAMS,usenatbib,usegraphicx]{mn2e}

\usepackage{amsmath}

\def\fermi{{\it Fermi\/}}
\def\auger{{\it Pierre Auger\/}}

\title[Correlating Fermi sources with UHECRs]{Correlating Fermi gamma-ray 
sources with ultra-high energy cosmic rays}
\author[Mirabal and Oya]{N. Mirabal$^{1,2}$\thanks{E-mail:
mirabal@gae.ucm.es} and I. Oya$^{2}$\\
$^{1}$Ram\'on y Cajal Fellow\\
$^{2}$ Dpto. de F\'isica At\'omica,
Molecular y Nuclear, Universidad Complutense de
Madrid, Spain\\}

\begin{document}

\date{}

\pagerange{\pageref{firstpage}--\pageref{lastpage}} \pubyear{2009}

\maketitle

\label{firstpage}

\begin{abstract}
The origin of ultra-high energy cosmic rays (UHECRs)  
is one of the enduring mysteries of high-energy 
astrophysics. To investigate this, we cross-correlate
the recently released \fermi\ Large Area Telescope First
Source Catalog (1FGL) with
the public sample of UHECRs made available by the
{\it Pierre Auger\/} collaboration. Of the 27 UHECRs in the sample, 
we find 12 events that arrived 
within $3.1\!^\circ$ of \fermi\ sources. However, we find similar 
or larger number of matches in 63 out of 100 artificial UHECR
samples constructed using positions randomly drawn from
the BATSE 4B catalog of gamma-ray bursts (GRBs) collected
from 1991 until 1996. Based on our analysis,
we find no evidence that UHECRs are 
associated with \fermi\ sources. We conclude with  
some remarks about the astrophysical origin of cosmic rays.
\end{abstract}


\section{Introduction}
Ultra-high energy cosmic rays (UHECRs) are energetic particles
($> 10^{19}$ eV)  that must originate in the 
most powerful particle accelerators in the Universe. These
extreme events have fascinated scientists from the time of 
their discovery
in 1962 \citep{linsley}. Since then, speculation about their origin
has flourished \citep{pierpaoli,ghise,cuesta}. 
Theoretically, active galactic nuclei (AGN) 
have long been favored as the best candidates for particle acceleration
to these extreme energies \citep{ginzburg,hillas}.  Unfortunately,
no firm astrophysical association has been 
established.

Possibly the most intriguing suggestion of an association between 
UHECRs and AGN was put forward by the {\it Pierre Auger\/} collaboration
\citep{abraham} that found a possible correlation between 
the arrival direction of 27 events collected in their Southern Observatory 
and the position of nearby ($\leq 75$ Mpc) AGN 
from the Ver\'on-Cetty catalog \citep{abraham2}. The
correlation remains but has weakened slightly with the inclusion of 
additional UHECR events (for a grand total of 58) collected between  
2004 January and 2009 March \citep{abraham3}. Nonetheless, 
questions remain about the likelihood of the
reported correlation \citep{abbasi2}. 

It is expected that if AGN are responsible for
UHECRs, these must be preferentially placed within a distance of 100
Mpc \citep{abraham2}.  Interactions with Cosmic Microwave Background
(CMB) photons should starve UHECRs arriving from larger distances
through the Greisen-Zatsepin-Kuzmin (GZK) effect
\citep{greisen,zatse}. In fact, a combination of recent observations
appear to corroborate the existence of a suppression of UHECRs above
$4 \times 10^{19}$ eV consistent with a GZK horizon
\citep{abbasi,abraham4}. It is important to note that
recent results from the {\it Pierre Auger\/}
  collaboration point to a transition to heavier composition with
  increasing energies \citep{unger}. If indeed iron nuclei 
are the dominant component of UHECRs, the large deviation
  angles expected for heavy nuclei  would render 
the detection of astrophysical counterparts even the nearest ones 
nearly impossible \citep{abraham5}.

Here, we present a cross-correlation study of UHECRs and 
the recently released \fermi\ Large Area Telescope First
Source Catalog (1FGL) in the 100 MeV to 100 GeV energy
range \citep{abdo}.
In contrast with previous studies, our cross-correlation 
analysis is unique in that the
\fermi\ catalog comprises a wide range of {\it bona fide}
particle accelerators with gamma-ray production above $100$ MeV
directed in the Earth's direction. 
The Large Area Telescope on board \fermi\ 
offers a major improvement in sensitivity over previous
GeV detectors \citep{atwood}. In its survey observation mode, the
  LAT observes the entire sky every 3 hours. For individual sources, 
\fermi\ provides nearly uniform sky coverage down to a photon
flux of $4 \times 10^{-10}$
cm$^{-2}$ s$^{-1}$ between 1 and 100 GeV,
except for sources at low 
Galactic latitude ($|b|\leq 10^\circ$) where the diffuse emission 
dominates \citep{abdo}. 
Throughout the paper, we assume an $H_0 = $71
km~s$^{-1}$~Mpc$^{-1}$, $\Omega_m = 0.27$, $\Omega_{\Lambda} = 0.73$
cosmological model.

\begin{figure}
\hfil
\includegraphics[width=3.1in,angle=0.]{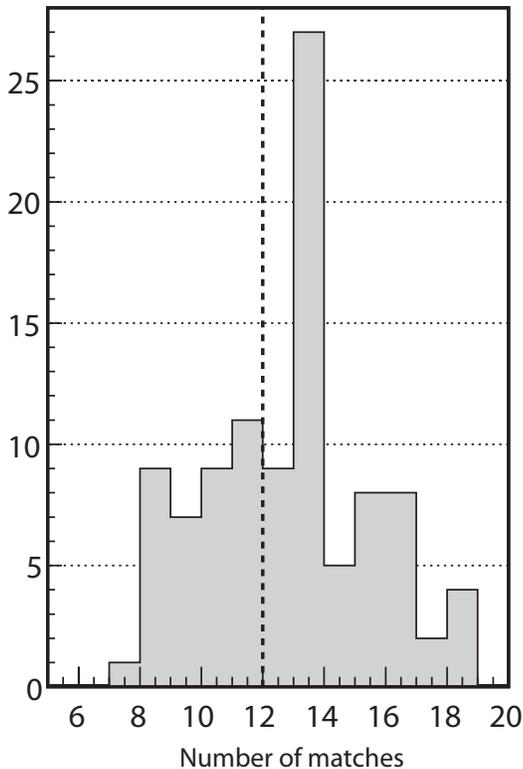}
\hfil
\caption{Distribution of total matches with the 1FGL catalog
for 100
random sample drawn from the BATSE 4B catalog. 
The vertical 
line indicates the number of matches between UHECRs detected
by the {\it Pierre Auger\/} Observatory and
the 1FGL catalog.
}
\label{figure1}
\end{figure}

\section{Cross-correlation of \auger\ UHECRs with \fermi\ sources}
As of 2007 November, 
the public catalog of UHECRs released by the {\it Pierre Auger\/} 
collaboration 
contains 27 events with energies above 55 EeV collected at their site in 
Malarg\"ue, Argentina \citep{abraham}. The \fermi\ LAT 
1FGL catalog consists of 1451 sources  characterized
in the 100 MeV--100 GeV energy range \citep{abdo}. 
The data were obtained in an all-sky scanning mode during 
2008 August--2009 July and represents the most extensive
map of the gamma-ray sky ($\geq 100$ MeV)  ever obtained. 
The entire catalog includes 689 blazars,
two starburst galaxies, two radio galaxies, 56 pulsars, 50 supernova 
remnants, and 630 unidentified sources \citep{abdo}. 

Since we are interested in testing  
for possible correlations of UHECRs with various classes
of gamma-ray emitters in the MeV-GeV energy range without
any {\it a priori}  assumption, we take advantage
of the complete 1FGL catalog without             
any redshift or type discrimination.

To test for a possible cross-correlation 
between UHECRs and \fermi\ sources in the 1FGL catalog, we checked whether 
the individual UHECR positions in the {\it Pierre Auger\/} sample
are clearly contained within the error circle of individual 
sources listed in the 1FGL catalog. Specifically, 
we count correlations whenever an UHECR event is within
a circle of $3.1\!^\circ$  radius around a particular \fermi\ source.
This radius is in line with the value predicted by 
conventional models of cosmic ray trajectories 
that consider the full effect of the Galactic 
magnetic field 
\citep{abraham5}. 
In order to avoid ``multiple'' counts,
we only consider one match per individual UHECR event.  

Within the 27 UHECR events, we find 12 matches with 1FGL sources.  To
  examine the likelihood of such correlation, we used the BATSE 4B
  gamma-ray burst (GRB) catalog \citep{paciesas} in place of 
a random generator
  of isotropic sky positions that allows us to generate 
artificial samples of simulated
UHECRs. The BATSE 4B catalog consists of 1637 GRB positions
localized by the BATSE instrument on board the Compton Gamma Ray
Observatory (CGRO) in the period between 1991 April 19 and 1996 August
29.  For our work, we restricted our analysis to events that would be
accessible from the {\it Pierre Auger\/} southern site 
at a declination $<$ $24.8\!^\circ$  
\citep{abraham}. In addition, we assume that the fraction of 
exposure is the same accross the declination range 
as stated in \citet{abraham5}. 
After the proper cuts were applied, we drew 100 random sets of 27
events from the resulting ``southern'' BATSE 4B catalog to match the
UHECR sample. We next proceeded to correlate each of the 100 random
sets with the 1451 \fermi\ sources in the same manner as with the
original {\it Pierre Auger\/} dataset.

Figure~\ref{figure1} shows the distribution of matches between the 
artificial samples of UHECRs constructed from the BATSE 4B catalog and 
the 1FGL catalog. In particular, 63$\%$ of the artificial samples 
have 12 or more matches consistent with \fermi\ positions.
Allowing for larger circles ($6^\circ$ and $8^\circ$ radii) only increases the
number of false positives when compared with the BATSE sample 
to 71$\%$ and 73$\%$ respectively.  
If the position of {\it Pierre Auger\/} events were strongly correlated with 
\fermi\ sources, we would expect a greater number of coincidences 
when compared
with the outcomes of 
randomly-generated artificial sets of UHECR events drawn
from the BATSE 4B GRB catalog. We find no evidence for such an association 
and subsequently cannot claim a positive cross-correlation 
between UHECRs and the 1FGL catalog. 

\section{Discussion and Conclusions}
In summary, we find no cross-correlation between UHECRs and 1FGL 
sources that cannot be reproduced by chance alignment. This differs  
from  the findings reported by \citet{abraham3}, using a different AGN
  sample \citep[see also][]{abraham2}. 
Examining the matches between the {\it Pierre Auger\/} events and the
1FGL catalog 
in closer detail, we notice that the sample includes one pulsar, two 
unidentified sources, and eight AGN. Four of the AGN have measured
well beyond the GZK horizon with 
redshifts as high as $z = 1.843$. The lowest redshifts correspond
to Centaurus A ($z = 0.002$) 
and NGC 4945 ($z = 0.002$), as we shall discuss later. 
The remaining two AGN associations have no 
spectroscopic redshifts. 

One could argue
that the aforementioned AGN without spectroscopic redshifts could potentially
form a parent population for UHECRs.
If it was, these AGN would most likely lie at distances closer than
100 Mpc \citep{abraham5}. Optical imaging studies of BL Lac reveal
that the host galaxies of nearby AGN with jets pointed in our
direction tend to cluster around a mean absolute magnitude in the $R$
band equivalent to $M_{R} = -22.8$ \citep{sba}.  For distances closer
than 100 Mpc, this would correspond to an apparent $R$-band magnitude
$R = 12.2$ or brighter.  A quick analysis of optical Digitized Sky
Survey images centered on the positions of the two matched AGN without
spectroscopic redshifts reveals no optical sources brighter than $R =
12.2$ at the derived AGN positions \citep{abdo2}.  Therefore, it is
unlikely that any of the paired UHECRs/AGN are located within the
hypothesized GZK horizon at 100 Mpc.

Interestingly, we do recover the two UHECRs previously
associated with the radio galaxy Centaurus A 
\citep{abraham5} and one UHECR consistent with the position of the
starburst/Seyfert 2 NGC 4945 \citep{moskalenko}. 
Both Centaurus A
(1FGL J1325.6$-$4300) and NGC 4945 (1FGL J1305.4$-$4928) are detected
by \fermi. However, NGC 4945 only stands out as one of the two Seyfert
2 galaxies in the 1FGL catalog rather than for its qualifications as a 
UHECR accelerator \citep{abdo2}. On the other hand, radio galaxies such
as Centaurus A could potentially account for the acceleration of some
UHECRs to energies $> 55$ EeV \citep{dermer}. As a result, it is
important to continue to explore the connection (if any) between
UHECRs and Centaurus A.  However, it is troubling that 24 UHECR events
detected by {\it Pierre Auger\/} remain without an apparent
\fermi\ gamma-ray counterpart within 100 Mpc (we dub these ``orphan''
UHECRs). 

\begin{figure}
\hfil
\includegraphics[width=3.1in,angle=0.]{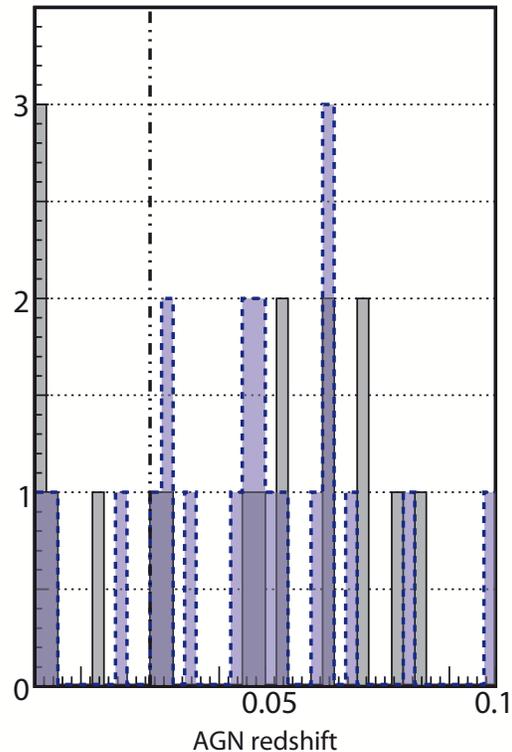}
\hfil
\caption{Redshift distribution of  \fermi\ AGN between $z=0$ and
$z=0.1$. Northern (decl $>$ $24.8\!^\circ$) and southern 
(decl $<$ $24.8\!^\circ$) samples are shown in blue (dashed line) and 
grey (solid line) respectively.
Only five identified \fermi\ sources have $z < 0.025$ and
are accessible to the {\it Pierre Auger\/} southern site.
The vertical line marks a distance of 100 Mpc. }
\label{figure2}
\end{figure}

In fact, there seems to be an apparent dearth of nearby AGN detected
in gamma rays at distances less than 100 Mpc ($z \leq 0.025$). To see
this more clearly, Figure~\ref{figure2} shows the distribution for
\fermi\ AGN in the redshift range $z=0-0.1$ \citep{abdo2}.  In total,
there are only five objects with $z \leq 0.025$ that would be
accessible to the {\it Pierre Auger\/} southern site namely NGC 253
($z= 0.001$), NGC 4945 ($z = 0.002$), Centaurus A ($z = 0.002$), M 87
($z = 0.004$), and ESO 323-G77 ($z = 0.015$).  NGC 4945 (being
generous) and Centaurus A could potentially account for 3 of the 27
UHECR events as discussed above.  However, we remark that there are no
additional identified \fermi\ sources in the 100 MeV-100 GeV energy
band to account for the bulk of UHECRs seen by {\it Pierre Auger\/} so
far.

It is possible that other AGN remain
unidentified in the 1FGL catalog,
as implied by the observed north-south anisotropy of
the associated sources, which might reflect the incompleteness of the
existing AGN catalogs in other bands \citep{abdo2}. However, 
the remainder is expected to lie in the lower end of the 
gamma-ray flux distribution, and therefore in principle 
not represent the most efficient 
particle accelerators. In addition, some AGN are known to be highly 
variable, therefore we cannot fully discard that  
some sources have escaped \fermi\ detection  and
possible correlation with UHECRs because of a low state.

Accordingly, one must admit one or all of the following possibilities:
1) if UHECRs are truly accelerated within the current sample of
  detected \fermi\ sources, their trajectories must experience
  deviations greater than $3.1\!^\circ$  from their actual
  astrophysical origin, making the identification of UHECRs nearly
  intractable. In particular, the absence of astrophysical counterparts could
be well justified if UHECRs are dominated by heavy nuclei including
iron since the deviation scales linearly the atomic number $Z$, 2) there could
potentially exist an undetected population of nearby AGN -- the so
called proton blazars?  \citep{mannheim} -- that only emit in a very
narrow band above 100 GeV to be discovered by future VHE experiments
such as the Cherenkov Telescope Array (CTA) or the Advanced Gamma
Imaging System (AGIS), and 3) the absence of obvious particle
accelerators in the MeV-GeV energy band leaves some room for the
exploration of additional forms of particle acceleration and even
exotic possibilities.

Trying to narrow down future observational directions, it is important
to search for possible timing/directional correlations of the arrival
direction of cosmic ray with AGN flares. Such exercise currently lacks
the dedicated sky monitoring that could potentially identify the
majority of AGN flares in real time. It would also require the rapid
release of UHECR detections (including position) in real time. The
existing fleet of MeV-TeV telescopes including {\it AGILE}, \fermi\,
H.E.S.S., MAGIC, and VERITAS could help trace strong flares in the
gamma-ray band. However, groundbreaking progress into the gamma-ray
variability domain might have to wait for the next generation of VHE
experiments. In the theoretical front, it is critical to improve
existing models of cosmic ray propagation. In particular, based on
current cross-correlation studies, the initial expectation that cosmic
ray trajectories at energies $> 10^{19}$ eV should be fairly rigid
does not appear to be so obvious.

\section*{Acknowledgments}
We thank all the members of Grupo de Altas Energ\'ias (GAE)
at the Universidad Complutense de Madrid for stimulating conversations during
our daily morning coffee. 
We also thank the anonymous referee for useful suggestions.
N.M. also acknowledges support from the Spanish Ministry of Science
and Innovation through a Ram\'on y Cajal fellowship.

\label{lastpage}
\end{document}